\documentclass[10pt]{article}

\usepackage{epsf,amsfonts,hyperref}
\renewcommand{\appendix}[1]{
    \addtocounter{section}{1}
    \setcounter{equation}{0}
    \renewcommand{\thesection}{\Alph{section}}
    \section*{Appendix \thesection\protect\indent #1}
    \addcontentsline{toc}{section}{Appendix \thesection\ \ \ #1}
}
\newcommand\encadremath[1]{\vbox{\hrule\hbox{\vrule\kern8pt
\vbox{\kern8pt \hbox{$\displaystyle #1$}\kern8pt}
\kern8pt\vrule}\hrule}}
\def\enca#1{\vbox{\hrule\hbox{
\vrule\kern8pt\vbox{\kern8pt \hbox{$\displaystyle #1$}
\kern8pt} \kern8pt\vrule}\hrule}}

\newcommand\figureframex[3]{
\begin{figure}[bth]
\hrule\hbox{\vrule\kern8pt
\vbox{\kern8pt \vbox{
\begin{center}
{\mbox{\epsfxsize=#1.truecm\epsfbox{#2}}}
\end{center}
\caption{#3}
}\kern8pt}
\kern8pt\vrule}\hrule
\end{figure}
}
\newcommand\figureframey[3]{
\begin{figure}[bth]
\hrule\hbox{\vrule\kern8pt
\vbox{\kern8pt \vbox{
\begin{center}
{\mbox{\epsfysize=#1.truecm\epsfbox{#2}}}
\end{center}
\caption{#3}
}\kern8pt}
\kern8pt\vrule}\hrule
\end{figure}
}

\renewcommand{\thesection}{\arabic{section}}

\makeatletter
\@addtoreset{equation}{section}
\makeatother
\newtheorem{theorem}{Theorem}[section]

\newtheorem{remark}{Remark}[section]
\newtheorem{proposition}{Proposition}[section]
\newtheorem{lemma}{Lemma}[section]
\newtheorem{corollary}{Corollary}[section]
\newtheorem{definition}{Definition}[section]
\def\br{\begin{remark}\rm\small}
\def\er{\end{remark}}
\def\bt{\begin{theorem}}
\def\et{\end{theorem}}
\def\bd{\begin{definition}}
\def\ed{\end{definition}}
\def\bp{\begin{proposition}}
\def\ep{\end{proposition}}
\def\bl{\begin{lemma}}
\def\el{\end{lemma}}
\def\bc{\begin{corollary}}
\def\ec{\end{corollary}}
\def\beaq{\begin{eqnarray}}
\def\eeaq{\end{eqnarray}}
\newcommand{\proof}[1]{{\noindent \bf proof:}\par
{#1} $\square$
\bigskip}

\newcommand{\eq}[1]{Eq.~(\ref{#1})}

\newcommand{\beq}{\begin{equation}}
\newcommand{\eeq}{\end{equation}}
\newcommand{\bea}{\begin{eqnarray}}
\newcommand{\eea}{\end{eqnarray}}

\renewcommand{\and}{{\qquad {\rm and} \qquad}}

\newcommand{\virg}{{\qquad , \qquad}}

 \newcommand{\Tr}{{\,\rm Tr}\:}

\newcommand{\Res}{\mathop{\,\rm Res\,}}

\newcommand{\ee}[1]{{{\rm e}^{#1}}}

\newcommand{\Pint}{{\int\kern -1.em -\kern-.25em}}

\renewcommand{\L}{\Lambda}

\newcommand{\ovl}{\overline}

\newcommand{\qbar}{\ovl{q}}

\begin{document}

\sloppy


\pagestyle{empty}
\addtolength{\baselineskip}{0.20\baselineskip}
\begin{center}
\vspace{26pt} {\large \bf {Symplectic invariants, Virasoro constraints and Givental decomposition}}
\newline
\vspace{26pt}

{\sl N.\ Orantin}\hspace*{0.05cm}\footnote{ E-mail: nicolas.orantin@cern.ch }\\
\vspace{6pt}
D\'epartement de math\'ematiques,\\
Chemin du cyclotron, 2\\
Universit\'e Catholique de Louvain,\\
1348 Louvain-la-Neuve, Belgium.\\
\end{center}

\vspace{20pt}
\begin{center}
{\bf Abstract}:
\end{center}

%

Following the works of Alexandrov, Mironov and Morozov, we show that the symplectic invariants of \cite{EOinvariants}
built from a given spectral curve
satisfy a set of Virasoro constraints associated to each pole of the differential form $ydx$ and each zero of $dx$. We then show that they
satisfy the same constraints as the partition function of the Matrix M-theory defined by Alexandrov, Mironov and
Morozov. The duality between the different matrix models of this theory is made clear as a special case of dualities
between symplectic invariants.
Indeed, a symplectic invariant admits two decomposition: as a product of Kontsevich integrals on the one hand, and as a product of 1 hermitian matrix integral on the other hand. These two decompositions can be though of as Givental formulae
for the KP tau functions.



\vspace{26pt} \pagestyle{plain} \setcounter{page}{1}


\section{Introduction and main results}

\subsection{Introduction: Matrix M-theory}

The theory of Hermitian random matrices is linked to many different fields in mathematics and physics by different very means
 (proved or conjectured) such as enumerative geometry, string theory or statistical physics.
But matrix models are not only dual to different theories, they are also dual from one another, e.g.
the hermitian one matrix model's partition function can lead to the Kontsevich integral by taking an appropriate
limit of the moduli of this model \cite{AK}. More generally, it seems that there exist dualities between many different matrix
models as it is pointed out in \cite{AMM1,AMM2,AMM3}. In this series of papers, the authors consider the possibility of the existence of
a random matrix equivalent of M-theory, i.e. they claim, and give evidences, that there should exist a general "M-theory" whose partition function
reduces to different kinds of matrix models in different patches of the moduli of this big theory\footnote{More precisely, this
partition function is defined as the string theory partition function in the sense of \cite{M}.}. Even though such
a general partition function was not explicitly built in these articles, they proposed to characterize it as the zero mode
of a differential operator defined on an associated algebraic curve: the spectral curve. This global operator
is shown to decompose as a sum of local Virasoro operators defined in the neighborhood of the singularities of the spectral curve.

Recent progresses in the resolution of hermitian random matrix models \cite{E1MM,ec1loopF,eyno,CEO,EOinvariants} have led to the definition of an infinite
set of numbers $F^{(h)}$ associated to an arbitrary algebraic curve whether this algebraic curve comes from the study
of a matrix model or not. More precisely, these works point out that the matrix models should not be studied only for themselves
but are just {\em one representation of the fundamental symplectic integrable invariants in some very particular cases}.
These more general partition functions always possess a topological expansion in a formal parameter $N$ which is identified
with the size of the matrices to be integrated in the matrix model representations. When the spectral curve is identified
as coming from a matrix model\footnote{There exists a generic
procedure to associate a spectral curve to a given matrix model. For details see \cite{EOinvariants,BA} for example.},
this partition function is equal to the partition function of the matrix model considered. We are thus able to build a partition
function associated to an arbitrary algebraic curve and which reduces to the partition function of matrix models for
some particular curves: it is a good candidate for this "matrix M-theory" partition function.

Nevertheless, the link between the approach of Alexandrov, Mironov and Morozov \cite{AMM2} is not that obvious and it is an
interesting problem to make it clear since both approaches can benefit from one another. This is precisely the aim of the present
paper: we show that both definitions coincide since the recursion relations of \cite{EOinvariants} reduce to the Virasoro
constraints of \cite{AMM2} when expressed in the right variables.

\subsection{Main results}

Let us consider an algebraic equation ${\cal{E}}(x,y)$\footnote{The symplectic invariants can be built from more general plane curve but we restrict the study of this paper to algebraic curves. Indeed, one studies the variations
of the symplectic invariants with respect to the holomorphic moduli of the spectral curve.} represented by a compact Riemann surface $\Sigma$ and two meromorphic
functions $x$ and $y$ on it such that
\beq
\forall p \in \Sigma \, , \; {\cal{E}}(x(p),y(p)) = 0
\eeq
and the one-form $ydx$ has poles $\alpha_i$ of degree $d_i+1$ and $dx$ has simple zeroes $a_i$. In the following,
we refer to this equation as the {\em spectral curve}.
 Following \cite{EOinvariants},
one defines the simplectic invariants $F^{(g)}({\cal{E}})$ associated to this equation and build the partition function
\beq
{\cal{Z}}({\cal{E}})= e^{-{\displaystyle \sum_g} N^{2-2g} F^{(g)}({\cal{E}})}.
\eeq
The purpose of this paper is to find a set of differential operators annihilating this partition function.

One first shows, in theorem \ref{thglobconstrpole}, that the "loop equations" satisfied by the correlation functions can be seen as Virasoro
constraints annihilating the partition functions:
\beq
\widehat{\cal{L}}(p) {\cal{Z}} = 0
\eeq
with
\beq
{\cal{L}}(p):= {1 \over N^2} :{\cal{J}}^2(p): +
\sum_i \oint_{\alpha_i} {  :{\cal{J}}^2(q):
\over (z_i(q)-z_i(p))dz_i(q)}
\eeq
where the current is defined at any point $p$ of the spectral curve by
\beq
{\cal{J}}(p):= N ydx(p) + {1 \over N} \partial_{B(.,p)}
\eeq
with $\partial_{B(.,p)}$ the loop insertion operator of \cite{EOinvariants}.
We then precise the link between the recursion relations of \cite{EOinvariants} and the Virasoro constraints of \cite{AMM2}.
Indeed, we show that the recursive definition of the correlation functions is nothing but saying that the action of a Virasoro operator
located at the branch points annihilates the partition function (see theorem \ref{thglobconstrbranch}):
\beq
\widehat{\cal{L}}(p) {\cal{Z}} = 0
\eeq
with
\beq
\widehat{\cal{L}}(p) := \sum_i \oint_{a_i} {dE_{q}^{(i)}(p)\over  (y(q)-y(\qbar))dx(q)} :{\cal{J}}(q){\cal{J}}(\qbar):.
\eeq

In section \ref{seclocalVirasoro}, we show that, in appropriate coordinates, these global Virasoro operators project to
local operators
\beq
{\cal{L}}(p) \to \sum_{j=-1}^{+\infty} {dz_i(p) \over z_i(p)^{j+1}} {L}_j^{(i)} \;\;\; \hbox{as} \;\;\; p \to \alpha_i
\eeq
with the {\em discrete Virasoro operators}
\beq
{L}_j^{(i)} = {1 \over N^2} \left( 2 j {\partial \over \partial t_{j,i}} +
\sum_{l=1}^{j-1} l (j-l) {\partial^2 \over \partial t_{j-l,i} \partial t_{l,i}}\right)
+ \sum_{k=1}^{d_i} (k+j) t_{j,i} {\partial \over \partial t_{k+j,i}}
\eeq
and a similar result for the projection of $\widehat{\cal{L}}(p)$ around the branch points.

Finally, these local Virasoro constraints allow us to decompose the partition function ${\cal{Z}}$ as a product of one
hermitian matrix integral  on the one side and a product of Kontsevich integrals on the other side (see theorems \ref{thdecompopole} and \ref{thdecompobranch}):
\beq
{\cal{Z}}({\cal{E}})= e^{\cal{U}} \prod_i {\cal{Z}}_H({\bf t_i}) = e^{\widehat{\cal{U}}} \prod_i {\cal{Z}}_K({\bf \tau_i})
\eeq
with two inter-twinning operators ${\cal{U}}$ and $\widehat{\cal{U}}$ linking the global structure of the spectrale curve
and the local behavior of $ydx$ at its poles and its zeroes encoded in the moduli $t_{j,i}$ and $\tau_i$ respectively.
These decomposition formulae were actually already derived by Givental in the study of the multi-component KP $\tau$-function \cite{Giv1,Giv2} and observed by Chekhov in the matrix model framework in \cite{Chekhov}.

\br
This last decomposition formula is proved only in the case of a genus 0 spectral curve. Indeed, if the spectral curve has higher
genus, there exists an ambiguity in the definition of the terms of the decomposition which
requires further investigations.
\er

\section{Symplectic invariants}

Let us first summarize how the symplectic invariants and correlation functions are defined in \cite{EOinvariants} and introduce
some useful notations.

\subsection{Algebraic geometry: definitions and notations}

In the following one considers an algebraic equation of degree $d_x+1$ in $x$ and $d_y+1$ in $y$
\beq
{\cal{E}}(x,y)=0
\eeq
referred to as {\em the spectral curve}. More precisely, the spectral curve is the triple $(\Sigma,x,y)$ where
$\Sigma$ is a compact Riemann surface and $x(p)$ and $y(p)$ two meromorphic functions on it such that
\beq
\forall p \in \Sigma \, , \; {\cal{E}}(x(p),y(p))=0.
\eeq
One also needs to equip this Riemann surface with a basis of canonical cycles $\left\{{\cal{A}}_i,{\cal{B}}_i\right\}_{i=1}^g$,
where $g$ is the genus of $\Sigma$.

\subsubsection{Branch points and sheeted structure}

Given a fixed value of $x$, the equation ${\cal{E}}(x,y)$ has generically $d_y+1$ distinct solutions in $y$. This
means that there exist $d_y+1$ distinct points $p^i \in \Sigma$, $i=0,\dots,d_y$, corresponding to the same value of $x$:
\beq
\forall (i,j)= 0,\dots,d_y \, , \; x(p^i) = x(p^j).
\eeq
This corresponds to saying that $\Sigma$ can be viewed as $d_y+1$ copies of the Riemann sphere, each corresponding to one
particular $p^i$: one calls such a copy of $\mathbb{CP}^1$, a $y$-sheet.
But there also exist particular points $a_i$, $i=1,\dots, \#bp$, where two pre-images of a given complex number $x(p)$ coincide:
\beq
\exists i \neq j \; , \; \; p^i=p^j .
\eeq
These points are called {\em $x$-branch points} since they correspond to loci where two $y$-sheets meet. They are
solution of the equation
\beq
dx(a_i)=0.
\eeq
In the following, we always suppose that all the branch points are simple, i.e. they are simple zeroes of $dx$\footnote{
One can deal with higher order zeroes by merging such simple zeroes. It is studied in section 8 of \cite{EOinvariants}.}.
This restriction implies that as a point $q$ approaches a branch point $a_i$, there exists a unique point $\qbar$
such that $x(q)=x(\qbar)$, $y(q)\neq y(\qbar)$ and $\qbar \to a_i$ as $q \to a_i$.
Remark that the application $q \to \qbar$ is not globally defined but only locally near each branch point\footnote{Nevertheless
it can be globally defined if the curve is hyperelliptic since it is the application which exchanges the $y$-sheets.}.

\subsubsection{Fundamental differentials}

We denote by $du_i(p)$ the $g$ holomorphic differentials on $\Sigma$ normalized on the ${\cal{A}}$-cycles:
\beq
\oint_{{\cal{A}}_i} du_j(p) = \delta_{ij}.
\eeq

The {\em Bergman Kernel} $B(p,q)$ is the unique bidifferential having only one pole in $p$ located at $p \to q$
such that
\beq
B(p,q) = {dz(p) dz(q) \over (z(p)-z(q))^2} + \hbox{regular} \;\; \hbox{as} \;\; p \to q
\eeq
and normalized by
\beq
\oint_{{\cal{A}}_i}  B(p,q) = 0.
\eeq

One also defines the {\em third Abelian differential}
\beq
dS_{p,p'}(q) = \int_p^{p'} B(q,.)
\eeq
which has two simple poles in $q \to p$ and $q \to p'$ with respective residues 1 and -1.

One especially needs a particular case of this differential when the integration path lies in the neighborhood of a branch
point $a_i$ linking one point $p$ to its conjugate $\overline{p}$:
\beq
dE_p^{(i)}(q):={1 \over 2} \int_{p}^{\overline{p}} B(q,.).
\eeq

\subsubsection{Moduli of the curve}

The main differential involved in the theory of the symplectic invariants is the 1-form $ydx$. First of
all, it has $\#_{poles}$ poles $\alpha_i$ of respective degrees $d_i$. It can thus be described by its behavior around these
poles
\beq
\forall i=1,\dots,\#_{poles} \, , \; ydx(p) \sim \sum_{k=1}^{d_i} k t_{k,i} \, z_i^k(p) \, dz_i(p) \;\; \hbox{as} \;\; p \to \alpha_i
\eeq
where $z_i(p)={1\over \xi_i(z)}$ is the inverse of a local variable in the neighborhood $\alpha_i$, i.e. it has a simple pole in $\alpha_i$. It is build as follows:
if $x$ is regular at $\alpha_i$, set $\xi_i(z)=x(z)-x(p)$, and if $x$ has a pole of degree $d$ at $\alpha_i$, set $\xi_i(z)=x(z)^{-1/d}$.
One also needs to precise its cycles integrals
\beq
\epsilon_i = {1 \over 2 i \pi} \oint_{\alpha_i} ydx.
\eeq
The coefficients $t_{k,i}$ are called the {\em moduli at the pole} $\alpha_i$ and the cycle integrals $\epsilon_i$ are
the {\em filling fractions}.

These moduli define totally the differential form $ydx$. Indeed, using the Riemann bilinear formula \cite{Farkas,Fay},
one can write it
\beq
ydx(p) = \sum_{i,k} k t_{k,i} B_{k,i}(p) + \sum_i t_{0,i} dS_{\alpha_i,o}(p) + 2 i \pi \sum_i \epsilon_i du_i(p)
\eeq
where
\beq
B_{k,i}(p) := - \Res_{q \to \alpha_i} B(p,q) z_i(q)^k .
\eeq

\subsection{Symplectic invariants, correlation function and free energy}

We now have everything in hand to define the central objects of this theory.

\subsubsection{Definitions}

Following \cite{EOinvariants}, let us define recursively the $k$ points, genus $h$ {\em correlation functions} $W_k^{(h)}(p_1,\dots,p_k)$ as
a $k$-form by
\bd
Correlation functions are defined by
\beq
W_{k+1}^{(h)}(p,{\bf p_K}):= \sum_i \Res_{q \to a_i} {dE_q^{(i)}(p) \over (y(q)-y(\qbar)) dx(q)} \left[
{\displaystyle \sum_{m=0}^h \sum'_{J \subset K}} W_{j+1}^{(m)}(q, {\bf p_J}) W_{k-j+1}^{(h-m)}(\qbar,{\bf p_{K \backslash J}})
+ W_{k+2}^{(h-1)}(q,\qbar,{\bf p_K}) \right]
\eeq
where ${\displaystyle \sum'}$ in the RHS means that we exclude the terms with $(m,J)=(0,\emptyset)$, and $(h,K)$.
\ed

We also define the {\em symplectic invariants}
\bd
For $h\geq 2$, the genus $h$ symplectic invariant is given by
\beq
{\cal{F}}^{(h)}:= {1 \over 2-2h} \sum_i \Res_{q\to a_i} \Phi(q) W_1^{(h)}(q)
\eeq
where $\Phi(q)$ is any primitive of $ydx$, whereas for $h=0,1$, they are given by
\beq
{\cal{F}}^{(1)}:= -{1 \over 2} \ln \left(\tau_{Bx}\right) - {1 \over 24} \ln \left(\prod_i y'(a_i)\right)
\eeq
where $\tau_{Bx}$ is the Bergmann $\tau$-function defined in \cite{EKK,KK}
and
\beq
{\cal{F}}^{(0)}:= {1 \over 2} \sum_i \Res_{a_i} V_i ydx + {1 \over 2} t_{0,i} \mu_{i} - {1 \over 4i\pi} \sum_i
\oint_{{\cal{A}}_i} ydx \oint_{{\cal{B}}_i} ydx
\eeq
where
\beq
V_i(p):= \Res_{q \to \alpha_i} y(q) dx(q) \ln\left(1 - {z_i(p)\over z_i(q)} \right)
\eeq
and
\beq
\mu_i :=\int_{\alpha_i}^o \left(ydx dV_i + t_{0,i} {dz_i \over z_i} \right) + V_i(o) -t_{0,i} \ln (z_i(o)).
\eeq

\ed

The correlation functions and free energies can be seen as the terms of the topological expansions of some
complete functions depending on an extra-variable $N$:
\bd
Let the complete correlation functions and free energies be defined by
\beq
{\cal{W}}_k(p_1,\dots,p_k):= \sum_{h=0}^\infty N^{2-2h-k} W_k^{(h)}(p_1,\dots,p_k)
\eeq
and
\beq
{\cal{F}}({\cal{E}}):= \sum_{h=0}^\infty F^{(2-2h)}({\cal{E}}).
\eeq
Let the partition function associated to the spectral curve ${\cal{E}}(x,y)$ be
\beq
{\cal{Z}}({\cal{E}}):= e^{-{\cal{F}}({\cal{E}})}.
\eeq

\ed

\subsubsection{Variation wrt the moduli of the spectral curve}

These free energies and thus the partition functions are functions of the moduli of the algebraic curve ${\cal{E}}$.
Among all the properties of these functions, it is interesting to note that their variations with respect to the moduli follow
a simple rule (see \cite{EOinvariants}):
\bt
When the one form $ydx$ changes infinitesimally to $ydx(p) + \epsilon \Omega(p)$ with
\beq
\Omega(p) = \int_{\partial \Omega} \L(q) \, B(p,q)
\eeq
for some function $\L$ and an integration contour ${\partial \Omega}$ away from the branch points, the correlation functions and free
energies change as follows
\beq
W_k^{(h)}({\bf p_K}) \to W_k^{(h)}({\bf p_K}) + \epsilon \int_{\partial \Omega} \L(q) \, W_{k+1}^{(h)}(q,{\bf p_K})
\eeq
and
\beq
F^{(h)} \to F^{(h)} + \epsilon \int_{\partial \Omega} \L(q) \, W_{1}^{(h)}(q).
\eeq

\et
Since the moduli of the curve are encoded in the one form $ydx$, one can extract from this theorem the variation
of the correlation functions and free energies wrt them (see \cite{EOinvariants} for details).

More precisely, the correlation functions themselves were built as the result of a particular variation of the spectral curve
changing $ydx(p)$ to $ydx(p) + \epsilon B(p,q)$. This variation is encoded in the so-called {\em loop insertion operator} $\partial_{B(.,p)}$
defined by
\beq
\partial_{B(.,p)} ydx(q) = B(p,q).
\eeq
This operator, depending on one point $p$ of the spectral curve, can be used to summarize the variations of all the moduli
of the spectral curve at once by looking at its Taylor expansion around a singularity of the spectral curve. This is the
basis of the arising of the Virasoro constraints studied in the forthcoming sections.

\section{Global and local Virasoro contraints}

In this section, we show that the partition function ${\cal{Z}}({\cal{E}})$ associated to a given algebraic curve ${\cal{E}}$ is the zero mode of two
operator valued meromorphic 1-forms on the considered algebraic curve: the {\em global Virasoro operators}. More precisely,
one has one differential operator associated to the moduli of the spectral curve at the poles of $ydx$ whereas the other one
involves the moduli at the $x$-branch points. Moreover the first operator is equivalent to the so-called loop equations
whereas the second one is equivalent to the recursive solution of \cite{EOinvariants} defining the correlation functions.

\subsection{Loop equations and global Virasoro constraints from the poles}

In \cite{EOinvariants}, the correlation functions associated to a given algebraic
equation were related to the variations of the free energy with respect to the moduli of this equation chosen to be
the coefficients of the Taylor expansion of the 1-form $ydx$ near its poles. In the present
section, we go further, showing that one can extract from this information a general differential equation mimicking
the loop equations derived in the context of the random matrix theory. Indeed, the correlation functions
can be shown to satisfy some similar "loop equations":

\bt
For any set of points $\{p,p_1,\dots,p_k\} \in \Sigma^{k+1}$, the complete correlation functions $W_k(p_1,p_2, \dots, p_k):= {\displaystyle \sum_g} N^{2-2g-k} W_k^{(g)}(p_1,\dots,p_k)$
satisfy the loop equations:
\beq
\sum_{l=0}^{k}   W_{l+1}(p,{\bf p_L}) W_{k-l+1}(p,{\bf p_{K \backslash L}}) + {1 \over N^2} W_{k+2}(p,p,{\bf p_K}) = P_{1,k}(p,{\bf p_K})dx(p)^2
\eeq
where the function
\beq
P_{1,k}(p,{\bf p_K}):= \sum_i \oint_{\alpha_i} {{\displaystyle \sum_{l=0}^{k}}   W_{l+1}(q,{\bf p_L}) W_{k-l+1}(q,{\bf p_{K \backslash L}}) + {1 \over N^2} W_{k+2}(q,q,{\bf p_K})
\over (z_i(p)-z_i(q))dx(q)}
\eeq
is a function of $p$ with poles only at the poles of $ydx$.
\et
\proof{
It follows from the loop equations derived in the simplest mixed case in \cite{symmetry}. Consider the
loop equations  (3-29) of \cite{symmetry} divided by $H_{0,0}(p,q)$ and take the residue as $q$ approaches the poles $\beta_i$ of $xdy$.
The function $P_{1,k}(p,{\bf p_K})$ is then equal to ${\displaystyle \Res_{q \to \beta_i}} {\widetilde{U}_{k,0}(x(p),q;{\bf p_K})dy(q)
\over H_{0,0}(p,q)}$. The formula comes directly from the pole decomposition of the function $P_{1,k}(p,{\bf p_K})$.

}

As it was pointed out in \cite{EOinvariants}, the correlation functions can be seen as variations of the free energy when one
changes the moduli of the spectral curve.
Remember that one can see the correlation functions as the result of the action of the loop insertion operator on the free energy:
\beq \label{difF1}
\partial_{B(.,q)} {\cal{F}} = W_1(q)
\eeq
and
\beq\label{difF2}
\partial_{B(.,p_1)} \partial_{B(.,p_2)} {\cal{F}} = W_2(p_1,p_2)
\eeq
where the free energy is the sum of all the genus contributions: ${\cal{F}}:= {\displaystyle \sum_g} N^{2-2g} F^{(g)}$. It is also
important to remark that this loop insertion operator is just build to deal with the variations of all the moduli at the poles
at once. Thanks to these properties
one can show that the partition function ${\cal{Z}}= e^{- {\cal{F}}}$ is the solution of a differential equation
involving the moduli of the spectral curve.

\bt\label{thglobconstrpole}
For any point $p \in \Sigma$, the partition function satisfies
\beq
{\cal{L}}(p) {\cal{Z}} = 0
\eeq
where one defines the {\em global Virasoro operator}\footnote{The name Virasoro operator is related to the projection of this
operator around the poles (cf section \ref{seclocalVirasoro}).}
\beq\label{globalconstraint}
{\cal{L}}(p):= {1 \over N^2} :{\cal{J}}^2(p): +
\sum_i \oint_{\alpha_i} {  :{\cal{J}}^2(q):
\over (z_i(q)-z_i(p))dx(q)}
\eeq
where the current is defined on any point of the spectral curve by
\beq
{\cal{J}}(p):= N ydx(p) + {1 \over N} \partial_{B(.,p)}.
\eeq

\et

\proof{ It directly follows from the properties \eq{difF1} and \eq{difF2} and the loop equation.
}

The partition function is thus the zero mode of a global operator defined as an operator valued meromorphic differential on
the spectral curve. This operator, labeled by a point on the spectral curve, is in fact used to summarize different
variations of the partition functions wrt to the moduli of the spectral curve on its poles, i.e. the moduli of the underlying theory.

\subsection{Recursive relations and global constraints from the branch points}

In the preceding section, we used a detour to build a global operator on the spectral curve annihilating the partition function
in order to work with the moduli introduced in \cite{EOinvariants}.
Indeed, we first built some loop equations satisfied by the correlation functions in order to build this global operator
thanks to its projection around the poles of $ydx$.  But we could choose to describe the 1-form $ydx$ by its behavior at
its zeroes instead of its poles, introducing moduli of the spectral curve as the coefficients of the taylor expansion of $ydx$
around the $x$-branch points.

Let us consider the definition of $W_1^{(g)}$ for any $g>0$:
\beq
W_1^{(g)}(p) = \sum_i \Res_{q \to a_i} {dE_q^{(i)}(p)\over (y(q)-y(\qbar))dx(q)} \left[\sum_{h=1}^{g-1} W_1^{(h)}(q)
W_1^{(g-h)}(\qbar) + W_2^{g-1}(q,\qbar)\right].
\eeq
One can remark that the left hand side can be included in the RHS by writing\footnote{For detailed computations, refer to
equation (4-12) of \cite{CEO}.}:
\beq
W_1^{(g)}(p) = \sum_i \Res_{q \to a_i} {dE_{q}^{(i)}(p)\over (y(q)-y(\qbar))dx(q)} \left(ydx(q) W_1^{(g)}(\qbar)
+ ydx(\qbar) W_1^{(g)}(q)\right).
\eeq
After summing over the genus $g$ and writing the correlation functions in terms of the variations of the partition function
one get
\bt\label{thglobconstrbranch}
For any point $p$ on the spectral curve, the partition function is a zero mode of the {\em global Virasoro
operator} $\widehat{\cal{L}}(p)$
\beq
\widehat{\cal{L}}(p) {\cal{Z}} = 0
\eeq
with
\beq\label{defglobbranch}
\widehat{\cal{L}}(p) := \sum_i \oint_{a_i} {dE_{q}^{(i)}(p)\over  (y(q)-y(\qbar))dx(q)} :{\cal{J}}(q){\cal{J}}(\qbar):.
\eeq
\et

We have thus built a second global Virasoro operator on the spectral curve. Note that the first one \eq{globalconstraint} was defined
in terms of an integral around the poles of $ydx$ while the new one is given in terms of a contour around the zeroes
of this differential form. Its is also interesting to note that they carry the same form thanks to the following lemma
\bl
The global Virasoro operator \eq{defglobbranch} can be written
\beq\begin{array}{rcl}
\widehat{\cal{L}}(p) &:=& - \sum_i \oint_{a_i} {dE_{q}^{(i)}(p)\over  (y(q)-y(\qbar))dx(q)} :{\cal{J}}^2(q):\cr
 &=& \sum_i \widehat{\cal{L}}_i(p)
\end{array}
\eeq
with
\beq
\widehat{\cal{L}}_i(p):=- \oint_{a_i} {dE_{q}^{(i)}(p)\over  (y(q)-y(\qbar))dx(q)} :{\cal{J}}^2(q): .
\eeq
\el

\proof{ The proof relies on the properties of the correlation functions as one changes sheets. It can be built
from the equations (4-5), (4-7), (4-14) and (4-15) in
 \cite{CEO}.}

\section{Local Virasoro constraints}\label{seclocalVirasoro}

We have now defined two global operators annihilating the partition function and involving all the moduli of the spectral curve.
Let us now project these constraints in the neighborhood of different singularities to make the link with the Virasoro algebra
clear, i.e. we look at this differential equation in different regime in the moduli space of the spectral curve\footnote{It
is interesting to see that moving the point in the spectral curve really corresponds
to selecting a regime in the moduli space of this curve.}.

\subsection{Virasoro constraint at the poles and Hermitian one matrix model representation}\label{sec1MM}

When the argument of the loop insertion operator approaches a pole of $ydx$, one can expand the latter in terms of the local variable $z_{i}$:
\beq\label{limitpole}
{\cal{J}}(p) \sim \sum_{j=1}^{d_i} {dz_i(p) \over z_{i}^{j+1}(p)} {\partial \over \partial t_{j,i}}+ \sum_j j t_{i,j} z_i^{j-1}(p)dz_i(p).
\eeq
We can then project the global constraint \eq{globalconstraint} in the neighborhood of this pole
and get:
\bt For any point $p$ in the neighborhood of a pole $\alpha_i$ of $ydx$:
\beq
{L}_-^{(i)}(p) {\cal{Z}} =0
\eeq
where the {\em local Virasoro operator} is defined as the loop operator
\beq
{L}_-^{(i)} := \oint_{\alpha_i} {1 \over (z_i(q)-z_i(p)) dz_i(q)} :\widehat{J}^{(i)}(q)^2:
\eeq
with the current
\beq
{J}^{(i)}(p):= \sum_{k\geq 0} \left[ {k t_{k,i} \over 2} z_i(p)^{k-1} dz_i(p) + {dz_i(p) \over z_i(p)^{k+1}} {\partial \over \partial t_{k,i}}\right].
\eeq
\et

It can be convenient to write this operator as
\beq
{L}_-^{(i)}(p) = \sum_{j=0}^{+\infty} {dz_i(p) \over z_i(p)^{j+1}} {L}_j^{(i)}
\eeq
with the {\em discrete Virasoro operators}
\beq
{L}_j^{(i)} = {1 \over N^2} \left( 2 j {\partial \over \partial t_{j,i}} +
\sum_{l=1}^{j-1} l (j-l) {\partial^2 \over \partial t_{j-l,i} \partial t_{l,i}}\right)
+ \sum_{k=1}^{d_i} (k+j) t_{j,i} {\partial \over \partial t_{k+j,i}}.
\eeq
It is easily checked that they indeed satisfy the commutation relations
\beq
\left[{L}_j^{(i)},{L}_l^{(k)}\right] = (j-l) {L}_{j+l} \delta_{i,k}.
\eeq

Thus, the symplectic invariants are $D$-modules in the sense of \cite{AMM1,AMM2,AMM3} as they are solution to some Virasoro constraints.
It is remarkable that one can associate one set of Virasoro constraints to each pole of $ydx$. One can actually
consider these different Virasoro algebra as local realizations of the global constraints imposed by the global Virasoro operator
\ref{globalconstraint}.

It is also interesting to note that the Virasoro constraints associated to a pole of the algebraic curve appears explicitly in the study of the Hermitian
one matrix model defined by the partition function:
\beq
{\cal{Z}}_H\left(\left\{t_k\right\}\right) := \int_{{\cal{H}}_N} dM e^{N {\displaystyle \sum_{k=0}^{d}} {t_k} \Tr M^{k}}
\eeq
where one integrates over $N \times N$ Hermitian matrices $M$.

Indeed, in this case, the partition function is the symplectic invariant built from the spectral curve:
\beq
{\cal{E}}_H(x,y):= y^2 - {\left( \sum_k k t_k x^{k-1}\right)^2} + P(x)
\eeq
where $P(x)$ is a polynomial of degree at most $d-1$. Thus, the function $y$ has two poles:
one simple pole noted $\infty_y$ and one pole of degree $d$ denoted by $\infty_x$. Moreover, $x$ has a simple pole at
$\infty_x$ and one can precise the behavior of $ydx$:
\beq
ydx(p) \sim \sum_k k t_k x^{k-1}(p) dx(p) \qquad \hbox{as} \qquad
p \to \infty_x.
\eeq
We are thus in the case described in this section with the $t_i$ of the decomposition in the neighborhood of the pole
given by the coefficients of the polynomial action in the matrix integral. Thus, one gets:

\bt
The partition function ${\cal{Z}}_H$ is solution
of the Virasoro constraints:
\beq
{L}_j^{(i)} {\cal{Z}}_{H}\left(\left\{t_k\right\}\right) =0 \qquad, \qquad \hbox{for} \;  j \geq 0.
\eeq
\et

Since the $t_k$'s involved in this equation are the only parameters of this model, these equation, and thus the neighborhood
of this unique pole of $ydx$, are sufficient to describe this model. This is why this representation of the free energy
is often used when one encounters this type of Virasoro constraints: typically, around a pole of $ydx$ (or $xdy$)
a D-module can be represented under this form.

It means that in this regime where one plugs in only the moduli at one pole, the partition function reduces to the one
of the hermitian 1 matrix model.

\subsection{Virasoro constraints at the branch points and Kontsevich integral}\label{secKontsevich}

One can also build such operators in the vicinity of the branch points $a_i$ giving rise to the Kontsevich kernel by
the gaussian case decomposition.

For this purpose, one has to expand the one-form $ydx$ in the neighborhood of its zeroes, i.e. the branch points, to emphasize the
moduli at the branch points. We thus have to introduce local parameterizations in the vicinity of the branch points
$a_i$. Since one considers only simple branch points, one has a natural parameter in the vicinity of a branch
point $a_i$:
\beq
\hat{z}_i(p) := {\sqrt{x(p)-x(a_i)}}.
\eeq
When $p \to a_i$:
\beq
{\hat{z}_i(p)} \sim y(p) -y(a_i)
\eeq
thus
\beq
y(p)dx(p) \sim 2 y(a_i) \hat{z}_i(p) d\hat{z}_i(p) + 2 \hat{z}_i^{2}(p) d\hat{z}_i(p).
\eeq

More precisely, let us write down the Taylor expansion of the 1-form $ydx$ in the neighborhood of the branch point
$a_i$ in terms of the local variable $z_i(p)$:
\beq\label{taylorbranch}
ydx(p) = \sum_{j=2}^{\infty} \tau_{j,i} \hat{z}_i(p)^{j-1} d\hat{z}_i(p)
\eeq
with
\beq
\tau_{j,i} :=\Res_{p \to a_i} y(p) dx(p) \hat{z}_i^{-j-1}(p).
\eeq

One can now blow up the spectral curve around this branch point, by expressing it in terms of the local coordinate
$\hat{z}_i$. The blown up spectral curve admits a rational parameterization:
\beq\label{localcurve}
\left\{\begin{array}{l}
\tilde{x}(z) = z^2 \cr
\tilde{y}(z) = {\displaystyle \sum_{k=2}^\infty} \tau_{k,i} z^{k-2}\cr
\end{array}
\right. .
\eeq
It was proved in \cite{EOinvariants} that the local behavior of the symplectic invariants around a critical point
is given by the symplectic invariants of the blown up spectral curve. Thus, in the vicinity
of the branch points $a_i$, the symplectic invariants reduce to those of \eq{localcurve} which are given by the topological
expansion of the Kontsevich integral defined as follows:
\cite{kontsevitch,IZ}
\beq
{\cal{Z}}_K(\tau_{k,i}):=\int dM\,\,\ee{-N\Tr ({M^3\over 3}-M (\L^2+\tau_1))} \virg \tau_1={1\over N}\Tr {1\over \L}
\eeq
where $\L$ is a deterministic external matrix defined by
\beq
\tau_{k,i} = {1\over N}\Tr \L^{-k}.
\eeq

This matrix integral is known to satisfy continuous Virasoro constraints in terms of the $\tau_{j,i}$:
\beq
\forall j\geq 2 \, , \;\widehat{L}_j^{(i)} {\cal{Z}}_K(\tau_{k,i}) = 0
\eeq
where the operator $\widehat{L}_j^{(i)}$ can be found in \cite{AMM1}

Hence, the partition function ${\cal{Z}}$ satisfies:
\beq
\forall j\geq 2 \, , \; \forall i \, , \;\widehat{L}_j^{(i)} {\cal{Z}} = 0.
\eeq

\section{Givental like decomposition}

From the first section, one knows that there exists two families of Virasoro operators defined in the neighborhood of the
poles and the zeros respectively annihilating the global partition function:
\beq\left\{\begin{array}{l}
\left[{\displaystyle \sum_i} {\cal{L}}_i(p) + {\cal{L}}(p) \right] {\cal{Z}} = 0\cr
{\displaystyle \sum_i} \widehat{\cal{L}}_i(p) {\cal{Z}} \cr
\end{array}
\right. .
\eeq

It means that one can decompose the partition function ${\cal{Z}}$ in two ways:
\begin{itemize}
\item it is the product of zeros-modes of the operators ${\cal{L}}_i(p)$, which are nothing but the partition functions
of the one matrix model studied in section \ref{sec1MM};

\item it is the product of the zero-modes of the operators $\widehat{\cal{L}}_i(p)$ which are nothing but Kontsevich integrals
studied in section \ref{secKontsevich}.

\end{itemize}

This means that the partition function ${\cal{Z}}$ can be decomposed as a product of 1 matrix model integrals or Kontsevich
integrals (which are KdV tau-functions) up to some conjugation operator mixing the local variables at the branch points and
poles of $ydx$. This reproduces the decomposition formulae discovered by Givental for multi-component KP tau functions \cite{Giv1,Giv2}.
Let us write these two types of decomposition explicitly.

Nevertheless, it must be noted that the Virasoro constraints and differential equations studied so far only involve one part
of the moduli of the spectral curve: they do not care about the filling fractions $\epsilon_i$. Thus, if the spectral curve
has non-vanishing genus, one should fix the dependence of both sides of the decomposition formula on these filling fractions
in order to get the right equality. This point is still under investigations and we consider in the following of this paper
that the spectral curve has genus 0.

\subsection{Decomposition of the global partition function in local partition functions}

Let us consider a prototype of Givental's like decomposition, i.e. a decomposition of the global Virasoro operator
as a product of local Virasoro operators in the neighborhood of a set of singular point $\xi_i$:
\beq
{\cal{L}}(p):= \sum_i \oint_{\xi_i} {dz_i(q) \over z_i(p) -z_i(q)} :{\cal{J}}(q)^2:
\eeq
with
\beq
{\cal{J}}(q)= N ydx(q) - {1 \over N} \partial_{B(.,q)}
\eeq
and the inverse of a local variable $z_i$ in the neighborhood of $\xi_i$.

Let us suppose that the global current ${\cal{J}}(p)$ converges to local currents ${\cal{J}}_i(p)$ as
$p \to \xi_i$ where
\beq
{\cal{J}}_i(p)= N \sum_k k t_{k,i} z_i^{k-1}(p) dz_i(p) + {1 \over N} { dz_i(p) \over z_i^{k+1}(p)} {\partial \over \partial t_{k,i}}
\eeq
for some fixed times $t_{k,i}$. We now decompose the zero mode ${\cal{Z}}$ of the global Virasoro operator
\beq
{\cal{L}}(p) {\cal{Z}} = 0
\eeq
into a product of the zero modes of the local operators:
\beq
{\cal{L}}_i(p) {\cal{Z}}_i = 0
\eeq
thanks to some conjugation operator ${\cal{U}}$:
\beq
{\cal{Z}} = e^{\cal{U}} \prod_i {\cal{Z}}_i.
\eeq

In order to compute the conjugation operator ${\cal{U}}$, one first identify the difference between the local currents and the
global one:
\beq
\Delta {\cal{J}}_i(p) = {\cal{J}}(p) - {\cal{J}}_i(p),
\eeq
since the intertwining operator is build to compensate this difference.

For this purpose, following \cite{AMM2}, let us define the bi-differential $f_{{\cal{O}},{\cal{O}}'}(p,p')$ associated to operators ${\cal{O}}(p)$
and ${\cal{O}}'(p)$:
\beq
f_{{\cal{O}},{\cal{O}}'}(p,p'):= {\cal{O}}(p) {\cal{O}}'(p') - :{\cal{O}}(p) {\cal{O}}'(p'):.
\eeq
Especially, one has
\beq
f_{{\cal{J}},{\cal{J}}}(p,p')= {1 \over N^2} B(p,p')
\eeq
and
\beq
f_{{\cal{J}}_i,{\cal{J}}_i}(p,p')= {1 \over N^2} {dz_i(p) dz_i(p') \over \left(z_i(p)-z_i(p')\right)^2}.
\eeq
One can compute explicitly
\beq
f_{{\cal{J}},{\cal{J}}}(p,p') - f_{{\cal{J}}_i,{\cal{J}}_i}(p,p') = \sum_{k,l} A_{k,l}^{(i)} z_i^k(p) z_i^l(p').
\eeq
On the other hand, one has
\bea
f_{{\cal{J}},{\cal{J}}}(p,p') - f_{{\cal{J}}_i,{\cal{J}}_i}(p,p') &=&
{\cal{J}}(p) {\cal{J}}(p') - :{\cal{J}}(p) {\cal{J}}(p'):
- {\cal{J}}_i(p) {\cal{J}}_i(p') + :{\cal{J}}(p) {\cal{J}}_i(p'): \cr
&=& \left[{\cal{J}}_i(p) + \Delta {\cal{J}}_i(p)\right] \left[{\cal{J}}_i(p') + \Delta {\cal{J}}_i(p')\right] -
:\left[{\cal{J}}_i(p) + \Delta {\cal{J}}_i(p)\right] \left[{\cal{J}}_i(p') + \Delta {\cal{J}}_i(p')\right]: \cr
 && - {\cal{J}}_i(p) {\cal{J}}_i(p') + :{\cal{J}}(p) {\cal{J}}_i(p'): \cr
&=& \Delta {\cal{J}}_i(p) {\cal{J}}_i(p') - :\Delta {\cal{J}}_i(p) {\cal{J}}_i(p'): .\cr
\eea
Looking for a generic solution of the form
\beq
\Delta {\cal{J}}_i(p) = \sum_{k,l} c_{k,l}^{(i)} z_i^k(p) {\partial \over \partial t_{l,i}},
\eeq
one gets
\beq
\Delta {\cal{J}}_i(p) = \sum_{k,l} A_{k,l}^{(i)} z_i^k(p) {\partial \over \partial t_{l,i}}.
\eeq
This can be written
\beq
\Delta {\cal{J}}_i(p) = \sum_j \oint_{\xi_j}  A^{(i,j)}(p,q) \Omega_j(q)
\eeq
where
\beq
A^{(i,j)}(p,p'):=f_{{\cal{J}},{\cal{J}}}(p,p') - f_{{\cal{J}}_i,{\cal{J}}_j}(p,p')
\eeq
and
\beq
\Omega_j(p):= N \sum_k t_{k,i} z_i^{k}(p) dz_i(p) - {1 \over N} { dz_i(p) \over k z_i^{k}(p)} {\partial \over \partial t_{k,i}}.
\eeq

Finally, the conjugation operator is constrained by
\beq
\forall i \, , \;  \Delta {\cal{J}}_i(p) = \left[{\cal{J}}_i(p),{\cal{U}}\right].
\eeq
This set of equations admits as solution
\beq\encadremath{
{\cal{U}}= \sum_{i,j} \oint_{\xi_j} \oint_{\xi_i} A^{(i,j)}(p,q) \Omega_j(q) \Omega_i(p).}
\eeq

Let us apply this analysis to the decomposition in one hermitian matrix model partition functions and Kontsevich integrals.

\subsection{Virasoro at poles: decomposition in 1 matrix models}

Let us first consider the case of the poles of $ydx$: $\left\{\xi_i\right\}:=\left\{\alpha_i\right\}$.
From \eq{limitpole}, one knows that the global current ${\cal{J}}(p)$ tends to the local current
${\cal{J}}_i(p)$ as $p \to \alpha_i$.
On the other hand, in section \ref{sec1MM}, we proved that the partition function of the one hermitian matrix model
\beq
{\cal{Z}}_H({\bf t}) = \int_{{\cal{H}}_N} dM e^{-N \sum_{k=0}^{d} {t_k} \Tr M^{k}}
\eeq
is a zero mode of the local Virasoro operator
\beq
{\cal{L}}_i(p)({\bf t}) {\cal{Z}}_H({\bf t}) = 0.
\eeq

The previous section implies then that
\bt\label{thdecompopole}
The global partition function can be decomposed in a product of one hermitian matrix integrals associated to the poles
$\alpha_i$ of the meromorphic form $ydx$
\beq\encadremath{
{\cal{Z}}({\bf t_1},{\bf t_2},\dots) = e^{\cal{U}} \prod_i {\cal{Z}}_H({\bf t_i}).}
\eeq
with the intertwining operator ${\cal{U}}$ defined by
\beq
{\cal{U}}:=\sum_{i,j} \oint_{\alpha_j} \oint_{\alpha_i} A^{(i,j)}(p,q) \Omega_j(q) \Omega_i(p)
\eeq
where
\bea
A^{(i,j)}(p,q)&:=& f_{{\cal{J}},{\cal{J}}}(p,p') - f_{{\cal{J}}_i,{\cal{J}}_j}(p,p')\cr
&=& B(p,q) - {dz_i(p) dz_j(q) \over (z_i(p)-z_j(q))^2} \cr
\eea
and
\beq
\Omega_i(p):= N \sum_k t_{k,i} z_i^{k}(p) dz_i(p) - {1 \over N} { dz_i(p) \over k z_i^{k}(p)} {\partial \over \partial t_{k,i}}.
\eeq

\et

\subsection{Virasoro at branch points: decomposition in Kontsevich integrals}

Let us now consider the case of the branch points: $\left\{\xi_i\right\}:=\left\{a_i\right\}$.

In section \ref{secKontsevich}, we proved that the Kontsevich integral
\beq
{\cal{Z}}_K:=\int dM\,\,\ee{-N\Tr ({M^3\over 3}-M (\L^2+\tau_1))} \virg \tau_1={1\over N}\Tr {1\over \L}
\eeq
is a zero mode of the local Virasoro operator
\beq
\widehat{\cal{L}}_i(p)({\bf \tau}) {\cal{Z}}_K({\bf \tau}) = 0.
\eeq

One thus has the decomposition formula
\bt\label{thdecompobranch}
The global partition function can be decomposed in a product of Kontsevich integrals associated to the branch points $a_i$:
\beq\encadremath{
{\cal{Z}}({\bf \tau_1},{\bf \tau_2},\dots) = e^{\widehat{\cal{U}}} \prod_i {\cal{Z}}_K({\bf \tau_i})}
\eeq
with the intertwining operator $\widehat{\cal{U}}$ defined by
\beq
\widehat{\cal{U}}:=\sum_{i,j} \oint_{a_j} \oint_{a_i} \widehat{A}^{(i,j)}(p,q) \widehat{\Omega}_j(q) \widehat{\Omega}_i(p)
\eeq
where
\bea
\widehat{A}^{(i,j)}(p,q)&:=& f_{{\cal{J}},{\cal{J}}}(p,p') - f_{\widehat{\cal{J}}_i,\widehat{\cal{J}}_j}(p,p')\cr
&=& B(p,q) - {d\hat{z}_i(p) d\hat{z}_j(q) \over (\hat{z}_i(p)-z\hat{z}_j(q))^2} \cr
\eea
and
\beq
\widehat{\Omega}_i(p):= N \sum_k \tau_{k,i} \hat{z}_i^{k}(p) d\hat{z}_i(p) - {1 \over N} { d\hat{z}_i(p) \over k \hat{z}_i^{k}(p)} {\partial \over \partial \tau_{k,i}}.
\eeq

\et

\section{Conclusion and perspectives}

In this paper, we investigated and made precise the link between the definition of the partition function of \cite{AMM2}
as a D-module and the recursive one of \cite{EOinvariants}. Indeed, we proved that the recursive definition of the correlation
functions of \cite{EOinvariants} are nothing but Virasoro constraints located at the branch points of the spectral curve whereas the
loop equations initially solved in the matrix models context can be seen as Virasoro constraints localized at the poles
of the one form $ydx$ on the spectral curve. This means that both approaches are totally equivalent as it was already pointed
out in \cite{MM} and thus the symplectic invariants are the string theory partition function studied in \cite{AMM2}.
We also completed the work of \cite{AMM2} by studying not only hyperelliptical curves but generic spectral curves. Moreover,
we have pointed out that the duality between the one hermitian matrix model and the Kontsevich integral just follows from
Virasoro constraints on the poles and on the branch points of the same spectral curve. Finally, one was led to decompose
the partition function as a product of random matrix integrals: 1 hermitian random matrix integrals at the poles and Kontsevich
integrals at the branch points. This gives a nice new representation of the symplectic invariants and a formalism complementary
to the approach of \cite{EOinvariants}.

It is interesting to note that these decomposition formulae were already derived by Givental to express the multi-component
KP tau-function as a product of KdV tau-functions (i.e. the Kontsevich integral). It was already mentioned in \cite{EOinvariants} that
the partition function build from the symplectic invariants could also be defined as the tau-function of an integrable model
(as it is proved for some matrix model's cases). The arising of the Givental decomposition formulae points also in this direction
and the formalism borrowed from \cite{AMM2} can be very useful to derive properly a Hirota equation which can be understood
as a defining property of the multi-component KP-tau function (e.g. see \cite{KP}), but this is left to a forth-coming work.

Moreover, as it was already remarked, we only focused in this paper on the moduli at the singularities of the spectral curve
and left the filling fractions aside. Nevertheless, it should be possible to study this other type of moduli in the same
way using the variation of the partition function wrt them. This step is fundamental if one wants to obtain Givental like decomposition
formulae for higher genus spectral curve and deserves further investigations.

Among the numerous possible applications of the symplectic invariants, let us mention two of them which can benefit directly
from this formalism and would merit further investigations. First, we only considered the "non-mixed
correlation functions"\footnote{This name is inherited from the study of multi-matrix model where these correlation functions
are observables which do not mix the different random matrices.} and the related loop equations. From this restricted set
of observables, one was able to extract Virasoro constraints which can be seen as a restriction of the more general
${\cal{W}}$-algebra constraints encountered for example in the 2-hermitian matrix models \cite{MMM2}. For this purpose,
it should be possible to express the mixed correlation functions of \cite{EOallmixed}, generalized away from the matrix models \cite{CEOmix},
as the result of the action of a differential operator on the partition function: these new operators should form some
${\cal{W}}$-algebra. Another aspect which should benefit investigations is the link between this formalism and Krichever-Novikov
like algebras \cite{KriNov}. Indeed, as it is pointed out in \cite{AMM2}, the currents ${\cal{J}}$ should satisfy some
commutation relations giving rise to a Krichever-Novikov algebra or, more precisely, it should give rise to its generalization
to arbitrary Riemann surface by Schlichenmaier \cite{Schlichenmaier}.

\vspace{2cm}

\noindent {\Large \bf Aknowledgements:}

I would like to thank Marc Adler, Mattia Cafasso, Bertrand Eynard and Pierre Van
Moerbeke for fruitful discuitions on the subject and particularly Leonid Chekhov for his comments on a first verion
of this paper.This work was supported by the ENRAGE European network MRTN-CT-2004-005616,
by the ENIGMA European network MRT-CT-2004-5652, by the French and Japanese governments through PAI SAKURA,
by the European Science Foundation through the MISGAM program and by the ANR project G\'{e}om\'{e}trie Int\'{e}grabilit\'{e}
en Physique Math\'{e}matique ANR-BLAN-0029-01.



\begin{thebibliography}{99}




\bibitem{KP} M.Adler, P.Van Moerbeke, P.Vanhaecke, ``Moment matrices and multi-component KP,
with application to random matrix theory'', math-ph/0612064.


\bibitem{AMM1} A.Alexandrov, A.Mironov and A.Morozov,
`` M-Theory of Matrix Models'', {\em Theor. Math. Phys.} {\bf 150} (2007) 179, hep-th/0605171.

\bibitem{AMM2} A.Alexandrov, A.Mironov and A.Morozov,
``Instantons and Merons in Matrix Models'', {\em Physica} {\bf D235} (2007) 126,
hep-th/0608228.

\bibitem{AMM3} A.Alexandrov, A.Mironov and A.Morozov, ``Partition function of matrix models as the first special functions
of string theory. I. Finite size Hermitean 1-matrix model, {\em Int. J. Mod. Phys.} {\bf A19} (2004) 4127,
{\em Theor. Math. Phys.} {\bf 142} (2005) 349, hep-th/0310113.

\bibitem{AK} J.Amnj{\o}rn, C.F.Kristjansen, ``From 1-matrix model to Kontsevich model'', {\em Mod. Phys. Lett.} {\bf A8}
(1993) 2875, hep-th/9307063.


\bibitem{CEOmix} L.Cantini, B.Eynard, N.Orantin, in preparation.


\bibitem{Chekhov} L.Chekhov, ``Matrix Models and Geometry of Moduli Spaces'', hep-th/9509001.


\bibitem{ec1loopF} L.Chekhov, B.Eynard,
``Hermitian matrix model free energy: Feynman graph technique for all genera'',
{\em J. High Energy Phys.} {\bf JHEP03} (2006) 014, hep-th/0504116.


\bibitem{CEO} L.Chekhov, B.Eynard and N.Orantin,
``Free energy topological expansion for the 2-matrix model'',
{\em J. High Energy Phys.} {\bf JHEP12} (2006) 053, math-ph/0603003.






\bibitem{E1MM} B.Eynard, ``Topological expansion for the 1-hermitian matrix model correlation functions'',
JHEP/024A/0904, hep-th/0407261.

\bibitem{EKK} B.Eynard, A.Kokotov, D.Korotkin, ``$1/N^2$ corrections to free energy in Hermitian two-matrix model'',
hep-th/0401166.

\bibitem{eyno} B.Eynard, N.Orantin,
``Topological expansion of the 2-matrix model correlation functions: diagrammatic rules for a residue formula'',
{\em J. High Energy Phys.} {\bf JHEP12}(2005)034, math-ph/0504058.


\bibitem{EOallmixed} B. Eynard, N. Orantin, ``Topological expansion and boundary conditions''
{\em JHEP} {\bf 06} (2008) 037, arXiv:0710.0223.



\bibitem{EOinvariants} B.Eynard, N.Orantin,
``Invariants of algebraic curves and topological expansion'', {\em Communication in Number Theory and Physics} {\bf vol.1 n°2},
math-ph/0702045.

\bibitem{symmetry} B.Eynard, N.Orantin, ``Topological expansion of mixed correlations in the hermitian 2 Matrix Model and x-y symmetry of the Fg invariants'',
{\em J. Phys. Math. Theor.}  {\bf A41} (2008) 015203,arXiv:0705.0958.

\bibitem{BA} B.Eynard, A.Prats Ferrer, ``Topological expansion of the chain of matrices'', arXiv:0805.1368.

\bibitem{Farkas} H.M. Farkas, I. Kra, ``Riemann surfaces'' 2nd edition, Springer Verlag, 1992.

\bibitem{Fay} J.D. Fay, ``Theta functions on Riemann surfaces'', Springer Verlag, 1973.

\bibitem{Giv1} A. Givental, ``Semisimple Frobenius structure at higher genus'', {\em Intern. Math.
Res. Notices} {\bf 23} (2001) 1265, math/0008067.

\bibitem{Giv2} A. Givental, ``$A_{n-1}$-singularities and nKdV hierarchies'', {\em Moscow. Math
Jour.} {\bf 3(2)} (2003) 475 , math.AG/0209205.

\bibitem{IZ} C.Itzykson, J.B.Zuber, ``Combinatorics of the Modular Group II : the Kontsevich integrals'',
{\em Int. J. Mod. Phys.} {\bf A7} (1992) 5661, hep-th/9201001.

\bibitem{KK} A.Kokotov, D.Korotkin, ``Bergmann tau-function on Hurwitz spaces and its applications'',
math-ph/0310008.


\bibitem{kontsevitch} M.Kontsevitch,
``Intersection theory on the moduli space of curves and the matrix Airy function'',
{\em Funk. Anal. Prilozh.} {\bf 25} (1991) 50-57;
Max-Planck Institut preprint MPI/91-47, MPI/91-77.


\bibitem{KriNov} I. Krichever, S.P. Novikov, ``Algebras of Virasoro type, Riemann surfaces and structure of the theory of
solitons'', {\em Funct. Anal. Appl.} {\bf 21} (1987) 126; {\em Funct. Anal. Appl.} {\bf 21} (1987) 294.



\bibitem{MMM2} A.Marshakov, A.Mironov, A.Morozov,
``From Virasoro constraints in Kontsevich's model to W-constraints in the 2-matrix model'',
{\em Mod. Phys. Lett.} {\bf A7} (1992) 1345, hep-th/9201010.



\bibitem{MM} A.Mironov, A.Morozov, ``Virasoro constraints for Kontsevich-Hurwitz partition function'', arXiv:0807.2843.

\bibitem{M} A. Morozov, ``String theory: What is it?'', {\em Sov. Phys. Usp.} {\bf 35} (1992) 671.



\bibitem{Schlichenmaier} M.Schlichenmaier, ``Differential operator algebras on compact Riemann surfaces'',
hep-th/9311036.

















\end{thebibliography}
\end{document}